\documentstyle[aps,epsf,multicol]{revtex}

\title{Error Thresholds on Dynamic Fitness-Landscapes}

\author{Martin Nilsson}

\address{Santa Fe Institute, 1399 Hyde Park Road, Santa Fe, New Mexico 87501 USA \\
Chalmers Tekniska H\"ogskola and G\"{o}tenborgs Universitet, G\"oteborg, Sweden {\tt martin@fy.chalmers.se}}

\author{Nigel Snoad}
\address{Santa Fe Institute, 1399 Hyde Park Road, Santa Fe, New Mexico 87501 USA \\
The Australian National University, ACT 0200, Australia {\tt nigel@santafe.edu}}
\date{24 February 1999}

\begin{document}
%\draft
\maketitle

\begin{abstract}

In this paper we investigate error-thresholds on dynamics fitness-landscapes. We show that there exists both lower and an upper threshold, representing limits to the copying fidelity of simple replicators. The lower bound can be expressed as a correction term to the error-threshold present on a static landscape. The upper error-threshold is a new limit that only exists on dynamic fitness-landscapes. We also show that for long genomes on highly dynamic fitness-landscapes there exists a lower bound on the selection pressure needed to enable effective selection of genomes with superior fitness independent of mutation rates, i.e., there are distinct limits to the evolutionary parameters in dynamic environments.

\end{abstract}

\begin{multicols}{2}
\narrowtext
Ever since Eigen's work on replicating molecules in 1971~\cite{Eigen71},
the concept 
of quasi-species has proven to be a very fruitful way
of modeling the fundamental behavior of evolution. A quasi-species 
is an equilibrium distribution of closely related gene sequences, localized 
around one or a few sequences with high fitness.  The combination of simplicity
and mathematical preciseness makes it possible to isolate the effects of
different fundamental parameters in the model. It also makes it possible
to capture some general phenomena in nature, such as the critical relation
between mutation rate and information transmission~\cite{Eigen71,Eigen77}.
The kinetics of these simple systems has been studied in great detail as the 
formulation has allowed many of the techniques of statistical physics to be 
applied to replicator and evolutionary systems. See  for 
instance~\cite{Eigen71,Eigen77,Schuster86,Schuster85,Leuthausser86,Tarazona92,Swetina88,NS89,EMcCS89,Bonhoeffer,Higgs94,AF96,AF97,AF98,BBW97}.  

The appearance in these models of an error-threshold (or error-catastrophy)
 as an upper bound on the mutation 
rate, above which no effective selection can occur, has important 
implications for biological systems. In particular it places limits on the 
maintainable 
amounts of genetic information~\cite{Eigen71,Eigen77,M-SS95} which puts strong 
restrictions on possible theories for the origins of life. 
It is interesting to note that some RNA-viruses seem to 
have evolved mutation rates that are close to the error-threshold 
\cite{Eigen77}\cite{M-SS95}. 

Studies of quasi-species until now have focused on static
fitness-landscapes. Many organisms in nature however live in a 
quickly changing environment \cite{VanValen73}. 
This is especially important for viruses and other microbial pathogens
 that must survive in a host with an highly dynamic immune system for which 
there only exist tight and temporary niches with high fitness (for the pathogen).

In this paper we investigate how the critical mutation rate of the error 
threshold
is affected by a dynamical fitness-landscape. We show how
the critical mutation rate is lowered by shifts of the fitness-peak.
An simple analytical expression for this critical copying fidelity
is also presented. It also turns out that if the selection pressure is too
small, the fitness-landscape moves too fast and the fitness encoding genome is too large, 
the population will lose the fitness-peak independent of mutation rate.
This shows the existence of regions in parameter space where no selection
can occur despite possibilities of adjusting copying-fidelity.

In brief a quasi-species consists of a population of
self-replicating genomes represented by a sequence of bases 
$s_k$, $\left( s_1 s_2 \cdots s_n \right)$. Hereafter we will assume binary 
bases $\{ 1, 0 \}$ and that all sequences have equal length $n$ though these 
restrictions are easily relaxed. 
Every genome is then given by a binary string $\left(011001 \cdots \right)$, 
which can be represented by an integer $k$ ($0 \leq k < 2^n$). 

To describe how mutations affect a population  we define $W_k ^l$ as
the probability that replication of genome $l$  gives genome
$k$ as offspring. For perfect copying accuracy, $W_k ^l$  equals
the identity matrix. Mutations however give rise to off diagonal
elements in $W_k ^l$. Since the genome length is fixed to $n$  
we will only consider point mutations, which conserve
the genome length.

We assume that the point mutation rate $p = 1 - q$ 
(where $q$ is the copying accuracy per base) is
constant in time and independent of position in the genome.
We can then write an explicit expression
for $W_k ^l$ in terms of the copying fidelity:

\begin{eqnarray}
    W_k ^l & = & p ^{h_{k l}} q ^{n - h_{k l}} = q ^n 
    \left( \frac{1-q}{q} \right) ^{h_{k l}} \label{eq1}
\end{eqnarray}
where $h_{k l}$ is the Hamming distance between genomes
$k$ and $l$, and $n$ is the genome length. The Hamming 
distance $h_{k l}$ is defined as the number of 
positions where genomes $k$ and $l$ differ.

The equations describing the dynamics of the population now take a 
relatively simple form. Let $x_k$
denote the relative concentration and $A_k$ the fitness of 
genome $k$. We then obtain the rate equations:

\begin{eqnarray}
     \dot{x} _k & = & \sum _l W_k ^l A_l x_l - e x_k 
             \label{eq2}
\end{eqnarray}
where $e  =  \sum _l A_l x_l$ and the dot denotes a time derivative.
The second term ensures the total normalization of the population 
(as $\sum _l x_l = 1$) so that $x_k$ describes relative concentrations.

To create a dynamic landscape we consider a single peaked fitness landscape 
\cite{Eigen77} whose peak moves, resulting in 
different optimal gene sequences at different times. 
Formally we can write $A_{k(t)} = \sigma$ and $A_l = 1$ $\forall l \neq k(t)$ where
the (changing) genome $k(t)$ describes how the peak moves through sequence space. 
If $k(t)$ is constant
in time the rate equation [Eq.~\ref{eq2}] corresponds to the classical (static) 
theory of quasi-species studied by Eigen and others.

We allow the peak in the fitness landscape to move to one of its closest 
neighbors (chosen ranomly). In this paper we assume that movements occur with a fixed 
frequency but one could also consider a probabilistic movement. 

The mutation matrix $W$ describes point mutations which occurr with equal 
probability independent of position in the
genome. This imposes a symmetry on the rate equations, dividing the 
relative concentrations into error classes $\Gamma_i$ described by their Hamming
distance $i$ from the master sequence ($\Gamma_0$). This reduces the effective dimension
of the sequence space from $2^n$ to $n+1$ thereby making the problem analytically 
tractable. The use of assymetric evolution operators (such as recombination) or 
fitness landscapes is obviously significantly more problematic and is the subject of ongoing work. When the fitness peak moves this landscape symmetry will be broken since one sequence in $\Gamma_1$ will be singled out 
as the new master sequence. This would only
affect the results we present below if the mean time between shifts in the fitness-landscape
was small --- as there would then be a substantial concentration of the old
master sequence present when the peak moves back into this error-class. We assume
the dynamics to be slow enough for this not to be a problem.  

\begin{figure}
\centering
\leavevmode
\epsfxsize = .75 \columnwidth
\epsfbox{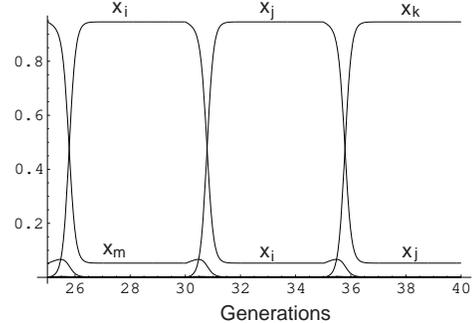}
\caption{This plot shows the population dynamics for gene
sequences of length $50$ when the fitness peak moves every
fifth generation ($\tau = 5$) and the height of the fitness peak $\sigma$
is $10$. The result is shown from $t = 25$ to remove initial transients.}
\label{timedynamics}
\end{figure}

Moving the
fitness peak then corresponds to applying the following co-ordinate 
transformation to the concentration vector:
\begin{eqnarray}
 R & = & \left( \begin{array}{ccccc} 0 & \frac{1}{n} & 0 &\cdots \\
                                  1 & 0 & \frac{2}{n} &  \cdots   \\
                                  0 & \frac{n-1}{n} & 0 & \cdots \\
                                  \vdots & \vdots & \vdots  & \ddots  
\end{array} \right)  \label{eq3}
\end{eqnarray}
To study the population dynamics we may divide the dynamics into cycles
 from time $0$ to $\tau$, where $\tau$ is a parameter
determining the number of generations between shifts of the fitness 
peak when the evolution proceeds as for a static landscape. We then apply 
the $R$
transformation to the concentration vector. The resulting concentration 
distribution is used as the
initial condition for the rate equations from time $\tau$ to $2 \tau$ 
and so on. These population dynamics [Eq.~\ref{eq2} and $R$] may be solved numerically as shown in 
Fig.~\ref{timedynamics} (after the initial 
transient) where $\tau = 5$, $\sigma = 10$, $q = 0.999$ and string-length $n=50$.

A simple approximation of the model presented above enables us to 
derive analytical expressions
for the error-thresholds on a dynamic fitness-landscape. 
Neglecting back mutations into the 
master-sequence, we can write the rate equation for the master-sequence 
on a static fitness-landscape as

\begin{eqnarray}
    \dot{x}_{mas} & = & Q \sigma x_{mas} - e x_{mas}
\label{approx}
\end{eqnarray}
where $Q = q^n$ is the copying fidelity of the whole genome and $e = \sigma x_{mas} + 1 - x_{mas}$.
 The asymptotic 
concentration of master-sequences is

\begin{eqnarray}
     x_{mas} \left( t \right) & \rightarrow  & \frac{Q \sigma - 1}{\sigma - 1} \mbox{    when }
 	t \rightarrow \infty
\label{asym}
\end{eqnarray}
This implies that the error-threshold on a static fitness-landscape
occurs when

\begin{eqnarray}
     Q^{stat} = \frac{1}{\sigma}
\end{eqnarray}
(see e.g.,~\cite{Eigen71,Eigen77}). This result is also intuitively clear 
since the superior fitness (and hence growth rate) of the master-sequence 
must compensate for the loss of $\Gamma_0$ individuals due to  mutations that 
occur during replication.

The intuitive picture of the error-threshold on a dynamic fitness-landscape
is different: what determines the critical mutation rate is whether
the master-sequence will have time to regrow between the shifts of
the fitness-peak. To find an analytical approximation for the 
error-treshold  we have to expand Eq.~(\ref{approx}) to include
the dynamics of error-class one as well as the master-sequence. This is
necessary since the fitness-peak moves into error-class one every $\tau$
time-steps. We can, however, make a large simplification by assuming 
the growth of the master-sequence to be in the exponential regime, i.e.,
that we can neglect the non-linearity in Eq.~(\ref{approx}). This is a good
approximation near the error-threshold as, for these values of $q$, 
the master-sequence
will not have time to approach any kind of equilibrium before the 
peak shifts again. We can thus write
an approximation of the rate equations for the master-sequence and
a representative member of error-class one:

\begin{eqnarray}
     \dot{x}_{mas} & = & \left( Q \sigma -1 \right) x_{mas} \nonumber \\
     \dot{x}_{1j} & = & \tilde{Q} \sigma x_{mas} + (Q-1) x_{1j}
\label{approx2}
\end{eqnarray}
where mutations into the member of error-class one are neglected and 
$\tilde{Q} = \left( 1-q \right)  q^{n-1}$ describes mutation from $x_{mas}$
into $x_{1j}$. We now assume $x_{1j} \left( 0 \right) = 0$, which is
a good approximation since $x_{1j}$ is (almost always) in $\Gamma_2$ 
before the shift. 
The solutions to Eq.~(\ref{approx2}) using this boundary condition
can be written as

\begin{eqnarray}
     x_{mas} \left( t \right) & = & x_{mas} \left( 0 \right) e^{\left( q^n \sigma -1 \right) t} \nonumber \\
     x_{1i } \left( t \right) & = & x_{mas} \left( 0 \right) \left( \frac{\left( e^{\left( q^n \sigma-1\right) t} - 
	e^{\left( q^n-1 \right) t} \right) \left( 1-q \right) \sigma}{\left( \sigma - 1 \right) q} \right) 
\label{solutions}
\end{eqnarray}

The shifting involves the move of the fitness peak
 to one of the sequences in error-class one 
at time $t = \tau$. The initial concentration of master-sequences at the 
beginning of a shift cycle is
therefore $x_{mas} \left( 0 \right) = x_1 \left( \tau \right)$. If the concentration of the master-sequence
after the shift is lower than immediately after the previous shift, i.e. 
$x_{mas} \left( 0 \right) > x_1 \left( \tau \right)$, the distribution of concentrations will
converge towards a uniform distribution. This is, in effect, a definition
of the error-threshold. A condition for effective selection
is then given by inserting $x_{mas} \left( 0 \right) < x_1 \left( \tau \right)$ into 
Eq.~(\ref{solutions}). We then derive a master-sequence growth parameter 

\begin{eqnarray}
\kappa \equiv \frac{ \left( e^{ \left( q^n \sigma - 1 \right) \tau} - 
	e^{ \left( q^n - 1 \right) \tau} \right) \left( 1 - q \right) \sigma}{\left( \sigma -1 \right) q} & > & 1
\label{error-eq}
\end{eqnarray}

\begin{figure}[h]
\centering
\leavevmode
\epsfxsize = .75 \columnwidth
\epsfbox{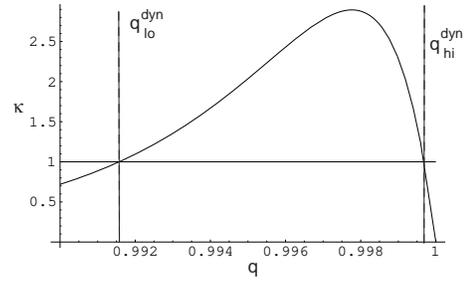}
\caption{The left hand side of Eq.~\ref{error-eq} is plotted as a 
function of the copying fidelity $q$. 
The genome length $n = 50$, $\tau = 2$ and $\sigma = 5$. The lower threshold 
is located at $q^{dyn}_{lo} = 0.988$
and the upper threshold at $q_{hi}^{dyn} = 0.9997$.} 
\label{curve}
\end{figure}

It is not possible to find
exact analytical solutions for the roots of Eq.~(\ref{error-eq}) and 
hence the error-thresholds. Fig.~\ref{curve} shows
the region where Eq.~(\ref{error-eq}) can be expected to hold. 
The figure also shows
the existence of two error-thresholds, $q_{lo}^{dyn}$ and $q_{hi}^{dyn}$ 
corresponding to the real roots of $\kappa=1$.
The lower threshold is a new version of the static error-threshold, with
a pertubation resulting from the movement of the fitness-landscape. 
The upper threshold 
is a new phenomenon that appears only on dynamic fitness-landscapes. 
Its existence is
intuitively clear --- if the mutation rate is very close to zero, 
there will not be enough individuals present on the new peak position 
when the shift occurrs to maintain a steady occupancy of the master 
sequence, i.e. the peak moves out from under the quasi-species 
and the population will not be
able to track shifts in the fitness-landscape.

Analytical approximations to the error-thresholds can be found by assuming different
dominant terms in the two different regions. To find the lower threshold $q_{lo}^{dyn}$ we
asssume $q^n$ to dominate the behavior. Solving for $q^n$ gives 

\begin{eqnarray}
      q^n & \approx & \frac{\tau - \ln 
	   \left( \frac{\sigma}{\sigma - 1} \cdot \frac{1 - q}{q} \right) }{\sigma \tau}
\label{eq-crit}
\end{eqnarray}
We can use Eq.~(\ref{eq-crit}) to find a first order correction in $\tau$ to
the static threshold by putting $q = \frac{1}{\sigma ^{1/n}}$
on the right hand side 

\begin{eqnarray}
     Q^{dyn}_{lo} & \approx & \frac{1}{\sigma} - \frac{\ln 
	   \left( \sigma ^{1/n}-1 \right) }{\tau \sigma }
\label{crit}
\end{eqnarray}
where we also made the approximation $\frac{\sigma}{\sigma-1} \approx 1$.
This is an expression for the lower error-threshold on a dynamic
fitness-landscape. Note that $Q_{lo}^{dyn} \rightarrow Q_{crit}$
when $\tau \rightarrow \infty$, i.e. we recover the stationary landscape limit. 

\begin{figure}[h]
\centering
\leavevmode
\epsfxsize = .75 \columnwidth
\epsfbox{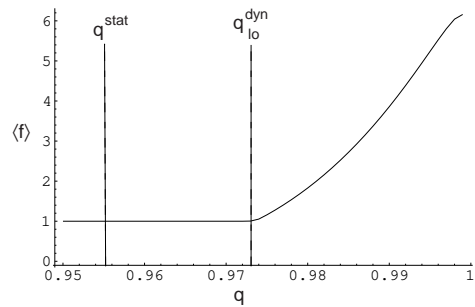}
\caption{The mean fitness is plotted as a function of the copying
fidelity per base $q$. The fitness peak moves every other generation
($\tau = 2$), the string-length $n$ is $50$ and the growth superority of the
master sequence $\sigma$ is $10$. The error-threshold occurs at 
the predicted value $q^{dyn}_{lo} = 0.973$. The 
static error-threshold is located at $q^{stat} = 0.955$.} 
\label{fit1}
\end{figure}

Fig.~\ref{fit1} shows the mean fitness of a population as a function
of the copying-fidelity. When $q$ is below $q_{lo}^{dyn}$, the 
concentration of master-sequences is approximately zero and the
mean fitness will therefore be $1$. The figure is based on numerical 
simulations of the full rate equations [Eq.~\ref{eq2}]. 
Note that the predicted 
value of $q_{lo}^{dyn}$ given by Eq.~ (\ref{crit}) is quite accurate.
Further comparisons to numerical solutions to the full dynamics are shown in table~\ref{table}.

Both the qualitative and quantitative dynamics of both error thresholds have
been verified by computer simulations using large populations to approximate
the deterministic dynamics.

\begin{figure}[h]
\centering
\leavevmode
\epsfxsize = .75 \columnwidth
\epsfbox{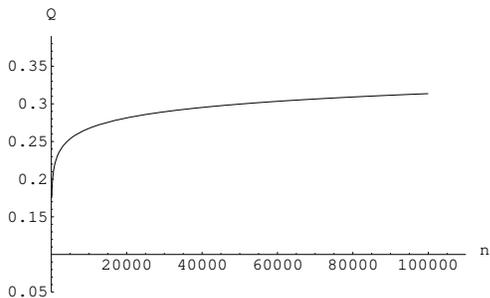}
\caption{The plot shows how the error-threshold $Q_{lo}^{dyn}$ given by
Eq.~(\ref{crit}) depends on the genome-length. The parameters describing the
fitness-landscape are fixed, $\sigma = 10$ and $\tau = 5$. The static error-threshold
is located co-incident with the x-axis at $Q^{stat} = 0.1$.} 
\label{copyfid}
\end{figure}

The critical copying fidelity $Q_{lo}^{dyn}$ 
depends on the genome-length. This is not 
surprising since the fitness-peak shifts into a specific member of $\Gamma_1$,
which consists of $n$ different gene-sequences. It is, however, a direct consequence
of the dynamic fitness-landscape since the static error-threshold is independent of
genome-length. This effect is demonstrated in Fig.~\ref{copyfid}, where $Q_{lo}^{dyn}$
versus the genome-length is plotted. The perturbation from the static error-threshold
increases with genome-length. The derivative is however decreasing and for reasonable
values of $\tau \gg 1$ and $\sigma \gg 1$ the static and dynamic error-threshold are of the same order 
of magnitude and show the same scaling behaviour.

\begin{table}[n]
\begin{tabular}{|c|c|c|c|c|c|} \hline 
  $\tau$ &  $\sigma$ & $n$ & $q_{threshold}$ & $q_{lo}^{dyn}$ & $q^{stat}$  \\ [0.3cm] \hline
  2  &  10   & 25 & 0.940           & 0.941            & 0.912      \\ [0.3cm] \hline
  2  &  10   & 50 & 0.973           & 0.973            & 0.955      \\ [0.3cm] \hline
  2  &  5    & 50 & 0.988           & 0.988            & 0.968      \\ [0.3cm] \hline
  5  &  10   & 50 & 0.963           & 0.964            & 0.955      \\ [0.3cm] \hline
\end{tabular}
\caption{The table shows results of numerical solutions of the error-threshold
compared to predicted values given by Eq.~(\ref{crit}) and the threshold for 
the corresponding static fitness-landscape.}
\label{table}
\end{table}
\nobreak

An analytical approximation to the new upper threshold can be found 
by assuming $q$ to be very close to $1$ and therefore 
the $\left( 1-q \right) $-term dominates the behaviour of Eq.~(\ref{error-eq}). 
Again assuming $\sigma \gg 1 $ and
putting $q^n = 1$, gives

\begin{eqnarray}
     q_{hi}^{dyn} & \approx & 1 - e^{-\left( \sigma -1 \right) \tau}
\label{upper}
\end{eqnarray}

Explicit numerical solutions of the full dynamics confirm that this threshold 
exsists and is predicted by Eq.~(\ref{upper}). 
For most values of $\sigma$ and $\tau$, $q_{hi}^{dyn}$ is very close to $1$ (e.g.
$\left( \sigma-1 \right) \tau = 50$ gives $10^{-22}$ as a  lower bound on the mutation rate per base pair). 
Finite population affects are however significant for the upper error-threshold. In
real biological populations this may be imposrtant. More detailed studies of these issues
are under preparation.

It is important to note that
$q_{hi}^{dyn}$ is independent of the genome-length. The total copying fidelity 
$Q_{hi}^{dyn} = \left( q_{hi}^{dyn} \right) ^n$ will then depend strongly on the genome-length. This
means that as the genome-length increases, the evolvable gap in between the two error-thresholds narrows.

On a static fitness-landscape it is always possible to find copying fidelities high enough
for evolution to be effective. It turns out that this is no longer the case for dynamic
fitness-landscapes. There exist regions in parameter-space (spanned by
$\sigma$, $\tau$ and $n$) where solutions to Eq.~(\ref{error-eq}) cease to exist. This happens when
the upper and lower error-thresholds coincide or, to put it differently, when the maximum (taken
over $q$) of the left hand side of Eq.~(\ref{error-eq}) become less than $1$. To find this convergence point it 
is better to search for a direct approximation of $q$ that maximizes the left hand 
side of Eq.~(\ref{error-eq}) as the approximations
for upper and lower error-thresholds given above become less accurate when they are close together, 
To do this we assume the leading behaviour is determined by the factor $e^{\left( q^n \sigma -1 \right) \tau}  \left( 1-q \right) $.
Taking the derivative of this expression and setting it to zero gives the equation
$q^{n-1} \left( 1-q \right) = \frac{1}{n \sigma \tau}$.
Assuming $q$ to be very close to $1$, and hence $q ^{n-1} \approx 1$ gives

\begin{table}[h]
\begin{tabular}{|c||c|c|c|c|c|} \hline 
  $\tau$ $\backslash$ $n$ & $50$  & $500$  & $5000$ & $50000$     & $10^9$ \\ [0.3cm] \hline \hline
  $1$         & $7.8$ & $10.4$  & $13.0$  & $15.5$ & $25.9$  \\ [0.3cm] \hline
  $10$        & $1.7$ & $2.0$  & $2.2$  & $2.4$    & $3.5$   \\ [0.3cm] \hline
  $50$        & $1.1$ & $1.2$  & $1.2$  & $1.3$    & $1.5$  \\ [0.3cm] \hline
\end{tabular}
\caption{The minimum selection pressure required for an infinite population to track the peak is listed for 
different values of the genome length $n$ and the number of generations between shifts of the fitness-peak 
$\tau$.}
\label{crit-points}
\end{table}
\nopagebreak

\begin{eqnarray}
     q_{max} & \approx & 1 - \frac{1}{\sigma \tau n}
\label{max}
\end{eqnarray}

This approximation for $q_{max}$ can be substituted into Eq.~(\ref{error-eq}). It is easy 
find points in phase space where this inequality starts to hold by fixing two parameters (e.g., $\tau$ and $n$) 
and then numerically solving for the third ($\sigma$). Table~\ref{crit-points} shows the minimal height
of the fitness-peak for different values of $\tau$ and $n$. The required selective pressure becomes
large for fast moving fitness-landscapes and large genome lengths.

In conclusion we have shown existence of, and derived analytic expressions for, 
two error-thresholds on a simple
dynamic fitness-landscape. The lower threshold is a perturbation of the well known 
error-catastrophy that exists a static 
fitness-landscape that accounts for the destabilizing effect of the changing environment. 
The existence of an upper bound on the copying fidelity is a new phenomenon, only existing in
dynamic environments. The presence of this upper bound results in the existence of critical 
regions of the landscape parameters ($\sigma$, $\tau$ and $n$) where the two thresholds
coincide (or cross) and threrefore no effective selection can occur. Thus dynamics 
landscapes have strong constraints on evolvability. 

We would like to thank Claes Andersson and Erik van Nimwegen for useful discussions. 
Thanks are also due to Mats Nordahl 
who has given valuable comments on the manuscript. Nigel Snoad and Martin Nilsson were supported by 
SFI core funding grants. N.S. would also like to acknowledge the support of Marc Feldman and the Center for Computational Genetics 
and Biological Modelling at Standford University while preparing this manuscript.

\bibliography{evolution}

\begin{thebibliography}{10}

\bibitem{Eigen71}
M.~Eigen.
\newblock Self-organization of matter and the evolution of biological
  macromolecules.
\newblock {\em Naturwissenschaften}, 58:465--523, 1971.

\bibitem{Eigen77}
M.~Eigen and P.~Schuster.
\newblock The hypercycle. {A} principle of natural self-organization. {P}art
  {A}: emergence of the hypercycle.
\newblock {\em Naturwissenschaften}, 64:541--565, 1977.

\bibitem{Schuster86}
P.~Schuster.
\newblock Dynamics of {M}olecular {E}volution.
\newblock {\em Physica D}, 16:100--119, 1986.

\bibitem{Schuster85}
P.~Schuster and K.~Sigmund.
\newblock Dynamics of {E}volutionary {O}ptimization.
\newblock {\em Ber. Bunsenges. Phys. Chem.}, 89:668--682, 1985.

\bibitem{Leuthausser86}
I.~Leuth\"ausser.
\newblock An exact correspondence between {E}igen's evolution model and a
  two-dimensional ising system.
\newblock {\em J. Chem. Phys.}, 84(3):1884--1885, 1986.

\bibitem{Tarazona92}
P.~Tarazona.
\newblock Error thresholds for molecular quasispecies as phase transitions:
  {F}rom simple landscapes to spin-glass models.
\newblock {\em Physical Review A}, 45(8):6038--6050, 1992.

\bibitem{Swetina88}
J.~Swetina and P.~Schuster.
\newblock Stationary {M}utant {D}istribution and {E}volutionary {O}ptimization.
\newblock {\em Bulletin of Mathematical Biology}, 50:635--660, 1988.

\bibitem{NS89}
M.~Nowak and P.~Schuster.
\newblock Error thresholds of replication in finite populations mutation
  frequencies and the onset of {M}uller's ratchet.
\newblock {\em J. theor. Biol.}, 137:375--395, 1989.

\bibitem{EMcCS89}
M.~Eigen, J.~McCaskill, and P.~Schuster.
\newblock The molecular quasispecies.
\newblock {\em Adv. Chem. Phys.}, 75:149--263, 1989.

\bibitem{Bonhoeffer}
L.S. Bonhoeffer and P.F. Stadler.
\newblock Error {T}hresholds on {C}orrelated {F}itness {L}andscapes.
\newblock {\em Journal of Theoretical Biology}, 164:359--372, 1993.

\bibitem{Higgs94}
P.G. Higgs.
\newblock Error thresholds and stationary mutant distributions in mulit-locus
  diploid genetics models.
\newblock {\em Genet. Res. Camb.}, 63:63--78, 1994.

\bibitem{AF96}
D.~Alves and J.F. Fontinari.
\newblock Population genetics approach to the quasispecies model.
\newblock {\em Phys. Rev. E}, 54(4):4048--4053, 1996.

\bibitem{AF97}
D.~Alves and J.F. Fontanari.
\newblock Error threshold in the evolution of diploid organisms.
\newblock {\em J. Phys. A.: Math. Gen.}, 30:2601--2607, 1997.

\bibitem{AF98}
D.~Alves and J.F. Fontanari.
\newblock Error thresholds in finite populations.
\newblock {\em Phys. Rev. E.}, 57(6):7008--7013, 1998.

\bibitem{BBW97}
E.~Baake, M.~Baake, and H.~Wagner.
\newblock Ising quantum chain is equivalent to a model of biological evolution.
\newblock {\em Phys. Rev. Lett.}, 78(3):559--562, 1997.

\bibitem{M-SS95}
J.~Maynard-Smith and E.~Szathm\'ary.
\newblock {\em The Major Transitions in Evolution}.
\newblock Oxford University Press: New York, 1995.

\bibitem{VanValen73}
L.~Van~Valen.
\newblock A new evolutionary law.
\newblock {\em Evol. Theory}, 1:1--30, 1973.

\end{thebibliography}

\bibliographystyle{unsrt}
\end{multicols}
\end{document}